\documentclass[reprint,amsmath,amssymb, aps,pre,floatfix,superscriptaddress]{revtex4-1}

\usepackage{graphicx} 
\usepackage{siunitx}
\usepackage[version=4]{mhchem}
\usepackage{comment}
\usepackage{xcolor}
\usepackage{tikz}
\usepackage{scalerel}
\usepackage{xifthen}
\usepackage{chngcntr}
\usepackage{xcolor}
\usetikzlibrary{svg.path}
\usepackage[colorlinks,allcolors=blue]{hyperref}

\newcommand{\figref}[2][{}]{Fig.\ \ref{#2}\ifthenelse{\isempty{#1}}{}{\,(#1)}}
\renewcommand{\vec}{\mathbf}

\definecolor{orcidlogocol}{HTML}{A6CE39}
\tikzset{
  orcidlogo/.pic={
    \fill[orcidlogocol] svg{M256,128c0,70.7-57.3,128-128,128C57.3,256,0,198.7,0,128C0,57.3,57.3,0,128,0C198.7,0,256,57.3,256,128z};
    \fill[white] svg{M86.3,186.2H70.9V79.1h15.4v48.4V186.2z}
                 svg{M108.9,79.1h41.6c39.6,0,57,28.3,57,53.6c0,27.5-21.5,53.6-56.8,53.6h-41.8V79.1z M124.3,172.4h24.5c34.9,0,42.9-26.5,42.9-39.7c0-21.5-13.7-39.7-43.7-39.7h-23.7V172.4z}
                 svg{M88.7,56.8c0,5.5-4.5,10.1-10.1,10.1c-5.6,0-10.1-4.6-10.1-10.1c0-5.6,4.5-10.1,10.1-10.1C84.2,46.7,88.7,51.3,88.7,56.8z};
  }
}

\newcommand\orcid[1]{\href{https://orcid.org/#1}{\mbox{\scalerel*{
\begin{tikzpicture}[yscale=-1,transform shape]
\pic{orcidlogo};
\end{tikzpicture}
}{|}}}}

\begin{document}

\setlength{\unitlength}{1cm}

\newcommand{\goeaffila}{Max Planck Institute for Dynamics and Self-Organization, Am Fa\ss{}berg 17, 37077 G\"ottingen, Germany}
\newcommand{\goeaffilb}{Institute for the Dynamics of Complex Systems, Georg August Universit\"at G\"ottingen, Germany}
\newcommand{\tokaffil}{Center for Exploratory Research, R\&D group, Hitachi Ltd., Higashi-Koigakubo 1-280, Kokubunji-shi, Tokyo 185-8601, Japan}
\newcommand{\twaffil}{Physics of Fluids Group, Max Planck Center for Complex Fluid Dynamics, MESA+ Institute and J. M. Burgers Center for Fluid Dynamics, University of Twente, PO Box 217,7500AE Enschede, Netherlands}
\newcommand{\hydaffil}{Department of Mechanical and Aerospace Engineering, Indian Institute of Technology Hyderabad, Kandi, Sangareddy, Telengana- 502285, India}

\title{Spontaneously rotating clusters of active droplets}

\author{Babak Vajdi Hokmabad~\orcid{0000-0001-5075-6357}}
\affiliation{\goeaffila}
\affiliation{\goeaffilb}
\author{Akinori Nishide}
\affiliation{\goeaffila}
\affiliation{\tokaffil}
 \author{Prashanth Ramesh~\orcid{0000-0001-6287-4107}}
\affiliation{\goeaffila}
\affiliation{\twaffil}

 \author{Corinna C. Maass~\orcid{0000-0001-6287-4107}}
 \email{corinna.maass@ds.mpg.de}%
\affiliation{\goeaffila}
\affiliation{\goeaffilb}
\affiliation{\twaffil}
\date{\today}%

\begin{abstract}
We report on the emergence of spontaneously rotating clusters in active emulsions. Ensembles of self-propelling droplets sediment and then  self-organise into planar, hexagonally ordered clusters which hover over the container bottom while spinning around the plane normal. This effect exists for symmetric and asymmetric arrangements of isotropic droplets and is therefore not caused by torques due to geometric asymmetries. We found, however, that individual droplets exhibit a helical swimming mode in a small window of intermediate activity in a force-free bulk medium.
We show that by forming an ordered cluster, the droplets cooperatively suppress their chaotic dynamics and turn the transient instability into a steady rotational state.
We analyse the collective rotational dynamics as a function of droplet activity and cluster size and further propose that the stable collective rotation in the cluster is caused by a cooperative coupling between the rotational modes of individual droplets in the cluster.

\end{abstract}

\maketitle
\section{Introduction}
Collective active matter systems comprise individual agents that constantly consume energy and generate mechanical force. These autonomous agents collectively form new functional materials through complex hierarchical self-organisation; a dynamic self-assembly that gives rise to new emergent properties and functionalities~\cite{snezhko2011_magnetic,aubret2018_targeted}. 

Suspensions of colloidal particles are promising platforms to study the principles of self-assembly and to use the obtained insights to design and engineer smart materials. They exhibit a variety of collective behaviours when they are driven out of equilibrium. Whether driven externally~\cite{sapozhnikov2003_dynamic,snezhko2011_magnetic,snezhko2009_self-assembled} or internally~\cite{palacci2013_living,kruger2016_dimensionality,theurkauff2012_dynamic}, the initially disordered suspensions can spontaneously form flocks and vortices~\cite{bricard2013_emergence,bricard2015_emergent} or self-assemble into ordered crystal-like structures in the vicinity of a solid boundary. Another widely studied mechanism is motility-induced phase separation (MIPS), where the particle velocity depends on the local concentration of its counterparts~\cite{buttinoni2013_dynamical, palacci2013_living,cates2015_motility-induced}. Usually, MIPS studies do not include hydrodynamic interactions. On the other hand, recent research has investigated collective behaviour in active matter driven mainly by hydrodynamic interactions~\cite{marchetti2013_hydrodynamics,ramaswamy2019_active,schwarzendahl2018_maximum,singh2016_universal}.
Squirmers confined to a lower boundary by gravity have shown a rich phase behaviour in numerical studies~\cite{kuhr2019_collective}.
Experimentally, flow-induced self-assembly into ordered two-dimensional crystals near a boundary has been reported in suspensions of \textit{Thiovulum majus} bacteria ~\cite{petroff2015_fast-moving}, in suspensions of starfish \textit{Patiria miniata} embryos~\cite{tan2021_development}, or in spermatozoa forming hexagonal vortex arrays~\cite{riedel2005_self-organized}. 
Flow-induced phase separation has also been observed in suspensions of artificial active particles such as bottom-heavy Janus colloids~\cite{palacci2013_living, shen2019_hydrodynamic,shen2019_gravity}, and active droplets~\cite{kruger2016_dimensionality,thutupalli2018_flow-induced}. For example, Singh and Adhikari ~\cite{singh2016_universal} showed that active hydrodynamic torques tend to destabilize the order inside 2D dynamic crystals. However, order can be restored via built-in asymmetries such as bottom-heaviness~\cite{palacci2013_living} or chirality~\cite{petroff2015_fast-moving}.

The self-propelling ability of particles with built-in asymmetries is easy to intuit -- imagine a Janus colloid converting fuel only on one side, leading to a propulsive chemophoretic front-to-back flow. 
Furthermore, asymmetric clusters of chemophoretic particles have been proposed to self-propel even if the individual particles are feature only diffusive chemokinetics~\cite{varma2018_clustering-induced}.

However, autophoretic motion has been shown to evolve even from initially isotropic geometries via dynamic instabilities that arise from spontaneously broken symmetries. Here, for a self-propelling sphere, the usual pathway towards motion is by the excitation of multipolar modes in the interfacial flow with increasing Péclet number, driven by a nonlinear coupling between chemical and hydrodynamic fields. These modes have been established both in theory ~\cite{michelin2013_spontaneous,morozov2019_nonlinear,morozov2020_adsorption} and in experimental realisations like self-propelling microdroplets~\cite{izri2014_self-propulsion,maass2016_swimming,hokmabad2021_emergence,suda2021_straight-to-curvilinear}. 

These dynamic modes on the individual scale can translate to complex cooperative dynamics in active ensembles~\cite{kruger2016_curling,thutupalli2018_flow-induced}: 
In the present study, we demonstrate how spontaneous rotational instabilities arising on the individual particle level carry through to synchronised cooperative rotation in aggregates.
Our model system is an active oil-in-water emulsion: close to a boundary, nematic droplets form ordered two-dimensional clusters mediated by hydrodynamic attraction and cooperatively swim upwards~\cite{kruger2016_dimensionality}. While investigating the assembly dynamics of isotropic droplets, we recently found that these clusters start to rotate spontaneously and persistently while hovering over the container bottom. To understand the mechanism behind the cooperative rotation, we studied the individual swimming behaviour in force-free bulk conditions and discovered a transient helical swimming mode.
However, our experiments show that this mode is limited to a very narrow activity range before the propulsion becomes chaotic. We hypothesise that forming an ordered cluster stabilises the chaotic dynamics of individual droplets, thereby extending the range of activity wherein rotational states emerge.

The paper is organised as follows: we will first discuss the rotational instabilities in individual droplets, then demonstrate how hydrodynamic interaction leads to aggregation and then study in detail under which conditions these aggregates stabilise and rotate. We show that the stability and dynamics of the clusters depend on droplet activity and cluster size and propose symmetry arguments to explain the respective rotation dynamics of liquid crystalline ordered and disordered ensembles.

\begin{figure}
  \centering
  \noindent\includegraphics[width=.6\columnwidth]{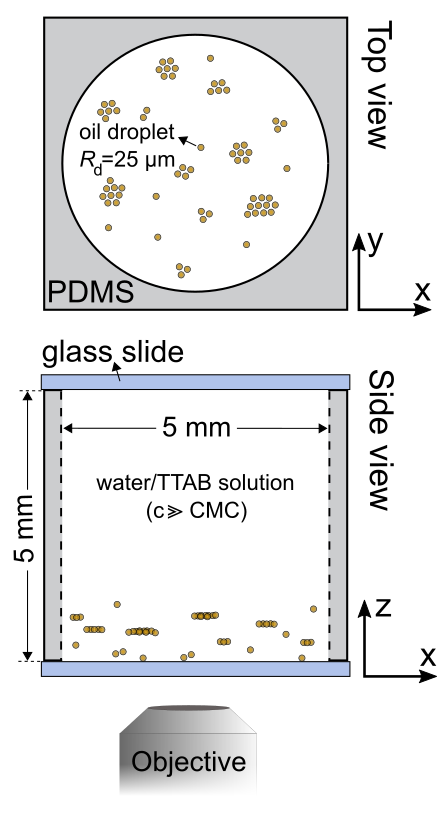}
  \caption{Schematic illustration of the experimental setup for bulk measurements, with an example sketch of a clustered sample. For quasi 2D experiments like PIV flow measurements, we used flat cells with a height corresponding to the particle diameter.
    \label{fig:setup} }
\end{figure}

\section{Broken rotational symmetries in individual droplets}
We begin with a description of the rotational instability in individual swimmers. The active emulsions we used comprise monodisperse droplets of CB15 oil (radius $R_d=\SI{25}{\um}$) in supramicellar aqueous solutions of TTAB surfactant at variable concentration ($c_{\rm TTAB}=6-20 \rm wt.\%$, where $\rm CMC=0.13 \rm wt.\%$). Over a typical time interval of several hours, these droplets dissolve slowly by micellar solubilization, which powers the self-propulsion. Thus, we can experiment at steady-state self-propulsion on a timeframe of several minutes to approximately half hours, with careful consideration of shrinkage effects.
The solubilization dynamics are characterised by a Péclet number $Pe$, quantifying the relative strength of the advective transport of surfactant molecules to their diffusion.

For our specific droplet model,  we can estimate $Pe$  via~\cite{hokmabad2021_emergence}:

\begin{equation} \label{eqn:pe2}
Pe=\frac{V_t R_d}{D} \approx \frac{18 \pi^2}{k_BT}  q_s r_s^2 \zeta R_d^2 \mu_i \left[\mu \left(\frac{2\mu+3 \zeta/R_d}{2 \mu+3}\right)\right].
\end{equation}
Here,  $V_t$ is the theoretical terminal droplet velocity in an external surfactant gradient~\cite{anderson1989_colloid,morozov2019_nonlinear}, $R_{\rm d}=25\,\mu$m the droplet radius, $D=\frac{k_BT}{6 \pi r_s \mu_o}$ the diffusion coefficient for the surfactant monomer (length scale $r_s \sim 10^{-10}$\,m), $q_s$ the isotropic interfacial surfactant consumption rate per area, $\zeta \sim 10$\,nm the characteristic length scale over which the surfactants interact with the droplet~\cite{anderson1989_colloid,izri2014_self-propulsion}. $\mu=\mu_o/\mu_i$, with $\mu_o$ and $\mu_i$ being the viscosities of the outer and inner phases.

 $Pe$ increases monotonically with droplet activity, which here is proportional to $q_s$. We control $Pe$ by varying $c_{\rm TTAB}$---an increase in $c_{\rm TTAB}$ results in higher $q_s$ (cf. Appendix~\ref{sec:clAppx}).

At supercritical $Pe\gtrsim 4$\cite{michelin2013_spontaneous}, the first, dipolar, mode in the interfacial surfactant coverage arises and leads to spontaneous motion, at typical speeds of $V\approx25\mu \rm m/\rm s$ in quasi-2D confinement~
\cite{herminghaus2014_interfacial,izri2014_self-propulsion,maass2016_swimming}.

\begin{figure*}
  \centering
  \noindent\includegraphics[width=\textwidth]{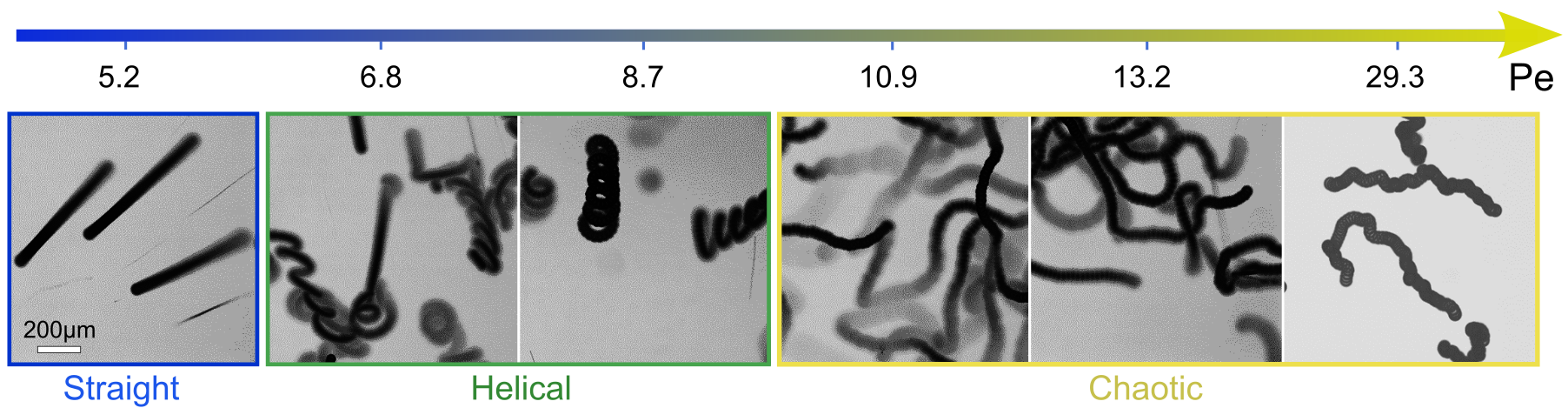}
  \caption{Dynamical behaviour of individual active droplets with varying $Pe$, transitioning from straight over helical to chaotic swimming. The $Pe$ range for each regime is shown by a different colour. 
    \label{fig:SingleDrop} }
\end{figure*}

To create force-free bulk conditions for 3D motion, we matched the density of the aqueous TTAB solution with that of the droplets by adding heavy water ($\rm D_2O$). 
In each experiment, we placed only a small number of droplets in the reservoir so that the inter-particle interactions are negligible and focused the microscope on a region well off the container boundaries. In the image series taken at increasing $Pe$ in \figref{fig:SingleDrop}, each image shows a multi-frame superposition  of individual droplets that visualises their trajectories. 
At low $Pe$, the droplets show a persistent and directed motion. At intermediate values ($Pe=$6.8 to 8.7), the droplets follow a helical trajectory. Since there is no global chirality, we conclude that the rotational symmetry is spontaneously broken.
At $Pe\geq10.9$, the droplets transition to increasingly chaotic swimming. 

We note that we have previously found a highly persistent and regular 'curling' (in 2D) or helical (in 3D) swimming mode in droplets composed of CB15's nematic isomer 5CB\cite{kruger2016_curling,suda2021_straight-to-curvilinear}. Here, the helical instability is due to a nematoelastic torque produced by a self-supporting asymmetric distortion of the director by hydrodynamic interactions. Oscillatory motion can also be generated by shear flow during the interaction of isotropic droplets with no-slip boundaries in Poiseuille flow~\cite{dey2021_oscillatory,zottl2012_nonlinear}.

However, none of these mechanisms apply to isotropic droplets in force-free bulk media. 
While a full understanding of this intermediate helical regime appears unfeasible without extensive numerical modelling, we can speculate about the possible origins based on previous findings for swimming in a quasi-2D environment~\cite{morozov2019_nonlinear,hokmabad2021_emergence,suda2021_straight-to-curvilinear}.

For our droplet system, we demonstrated that with increasing $Pe$ the propulsion transitions from steady and persistent to unsteady and chaotic due to non-linear interactions between flow and chemical fields~\cite{hokmabad2021_emergence}. 

In 2D, for intermediate $Pe$ values,  higher hydrodynamic modes are excited momentarily, generating a flow with quadrupolar symmetry. Such flow results in a metastable accumulation of oil-filled micelles (``spent fuel'') at the anterior stagnation point of the droplet. The formation of these accumulates and their advection to the droplet posterior causes quasi-periodic reorientations and a meandering trajectory. 

While the dynamics in a Hele-Shaw cell are confined to two dimensions, in a 3D bulk medium the droplet can reorient within the entire angular space, allowing for more complex modes of symmetry breaking which might lead to the more regular helical trajectories shown in \figref{fig:SingleDrop}. 

\section{Attractive interaction between droplets}

On the way towards cluster assembly, we now proceed to the mutual attraction between two droplets.

To experimentally investigate the droplet hydrodynamics, we measured to measure the flow field in the aqueous swimming medium generated by a droplet in a Hele-Shaw cell (height $h\approx 2R_{\rm d}$) by particle image velocimetry (PIV). To model the droplet flow field in confinement, we used an expansion ansatz corresponding to the Brinkman squirmer model~\cite{jin2021_collective}, (see Appendix \ref{app_Brinkman} for details). The first two modes of the Brinkman solution were then fitted to the PIV velocity field to determine the coefficients $b_1$ and $b_2$.

\figref[a]{fig:interaction} shows the flow field generated by the droplet which is in good agreement with the field as calculated using the
Brinkman squirmer model (\figref[b]{fig:interaction}). We extracted the tangential component of the velocity vector $u_{\theta}$ at the interface from the theoretical fit (\figref[c]{fig:interaction}) and observed that $u_{\theta}$ does not peak at the equator but closer to the rear stagnation point. Correspondingly, the squirmer parameter  $b_2/b_1=-0.692$   is consistent with a weak ``pusher'' flow in the far-field (see also numerical literature, as in  \cite{izri2014_self-propulsion}  and \cite{michelin2013_spontaneous}). We can further assume that the Brinkman squirmer in 2D confinement will correspond to a qualitatively similar classical squirmer in 3D.

Thutupalli et al.~\cite{thutupalli2018_flow-induced} attributed this pusher-type asymmetry to the balance between viscous and nematic
stresses in a liquid crystal droplet. 
However, since we observe a similar asymmetry in isotropic droplets,  nematic anisotropy cannot be the determining factor, and the asymmetry is more likely due the complex transport of surfactant molecules in the external phase. 

To investigate attractive interactions between droplets, we study two droplets in a Hele-Shaw cell. As shown in ~\figref[d,e]{fig:interaction} and Movie S1, the two droplets attract each other, and continue to swim  side-by-side for an extended period. This attraction can be rationalised by the hydrodynamic interaction between two pusher-type squirmers. 
The flow profile of a single pusher features an influx around the droplet equator~\cite{blake1971_spherical}. 
 Once two droplets swim on a converging course, this influx causes an attractive force~\cite{thutupalli2018_flow-induced}, such that they reorient each other to swim in parallel~\cite{lauga2009_hydrodynamics}. In our experiments, it is noteworthy that the droplets never come in contact and the stable parallel swimming happens at a finite interparticle distance larger than $2R_{\rm d}$. This is probably due to the overlap of the two chemorepulsive clouds of oil-filled micelles around the droplets\cite{moerman2017_solute-mediated}, in which case the equilibrium distance would be determined by the interplay of hydrodynamic attraction and phoretic repulsion.

\begin{figure*}[ht]
  \centering
  \noindent\includegraphics[width=0.7\textwidth]{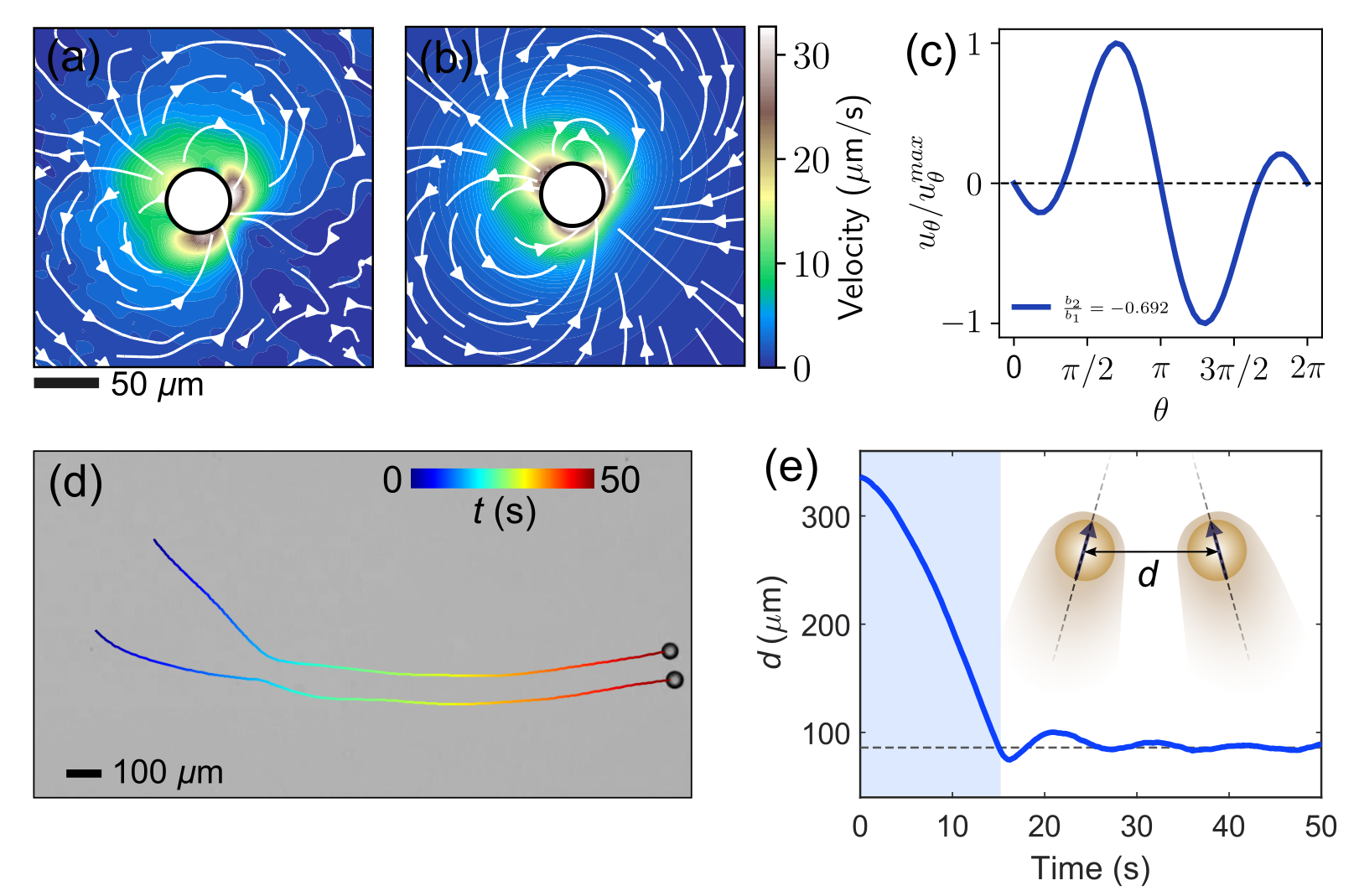}
  \caption{(a) Mean flow field generated by a single freely swimming droplet inside a Hele-Shaw cell. The colourmap shows the speed. (b) Theoretical flow field obtained from fitting the Brinkman solution to the experimental flow field. (c) Tangential component of the velocity $u_{\theta}$ at the vicinity of the interface obtained from the theoretical model. (d) The interaction between two active droplets.
  We note that the droplets in this dataset had shrunk somewhat due to solubilization, to $R<\SI{25}{\um}$, before we started recording.
   (e) Variation of the separation distance $d$ between the two droplets in time. The shaded region corresponds to the long-range hydrodynamic interactions leading to the formation of a droplet pair.
    \label{fig:interaction} }
\end{figure*}

\section{Formation of spontaneously-rotating clusters}
In a 3D system under gravity with higher number densities, the attractive hydrodynamic interactions discussed in the previous section lead to the flow-induced self-assembly of planar hovering clusters~\cite{kruger2016_dimensionality,thutupalli2018_flow-induced}.
To study the formation of such clusters, we filled a deep closed reservoir (5 mm in diameter and 5 mm in depth) with an active emulsion of monodisperse droplets. Due to their higher mass density, the droplets gradually sediment to the bottom of the reservoir. There, they start to interact with each other and autonomously assemble into planar clusters (Movie S2). 
This process takes place in four stages (\figref[a-c]{fig:clustering}): (i) \textit{Sedimentation}, due to gravity, (ii) \textit{Aggregation}, due to the mutual hydrodynamic attractive forces, (iii) \textit{Cluster formation}, during which the droplets form hexagonally-ordered crystal-like structures and all droplet reorient to swim upwards against gravity, and (iv) \textit{Rising and Rotation}, where the droplets cooperatively start to hover and rise, while spinning around the cluster axis. 

To illustrate the phenomenon, we tracked the $xy$ projection of the motion of seven individual droplets that eventually form a rotationally symmetric cluster. We have plotted the individual droplet trajectories centred around the coordinate of the middle droplet. As shown in \figref[d]{fig:clustering} and Movie S3, the droplets initially move about randomly. 
However, after aggregation, both the ordered cluster configuration and the rotational dynamics remain steady over long times. 

All of the individual droplets in a cluster reorient to swim upwards against gravity, thereby generating a global convective flow which makes the cluster hover (\figref[b]{fig:clustering})~\cite{kruger2016_dimensionality}.
Here, the in-plane configuration of the cluster is stabilised by the hydrodynamic interaction with the cell bottom~\cite{ruhle2018_gravity-induced}.
Using the calibrated z-drive of the stage of an Olympus IX-83 microscope at $40\times$ magnification, we measured the hovering height of clusters with different sizes. \figref[e]{fig:clustering} demonstrates the variation of cluster height in time. Regardless of the cluster size (number of droplets $N$), they all tend to gradually rise with a speed on the order of $1 \mu \rm m/s$. 
We attribute the rising of the clusters to the fact that the droplets dissolve and their mass decreases accordingly. This motion continues up to a height of several hundred micrometers, where the stabilising effect of the cell bottom is so weak that the cluster becomes susceptible to perturbations and eventually  disintegrates. 

\figref[a-f]{fig:flowfields} show the flow fields generated by an non-rotating (top row) and a rotating (bottom) cluster in $xy$ planes at different heights $z$. Irrespective of rotation, the radial components of the velocity field $v_r$ show similar behaviour: outward flow ($v_r>0$) underneath the cluster and inward flow above the cluster ($v_r<0$). However, the tangential flow $v_{\theta}$ only emerges at rotating cluster and is strongest in the cluster plane. The visualisation of the convective flows within the droplets in \figref[g,h]{fig:flowfields} and Movie S4 demonstrate the difference between rotating and non-rotating clusters although still the top-down flow is dominant.

\begin{figure}
  \centering
  \noindent\includegraphics[width=0.5\textwidth]{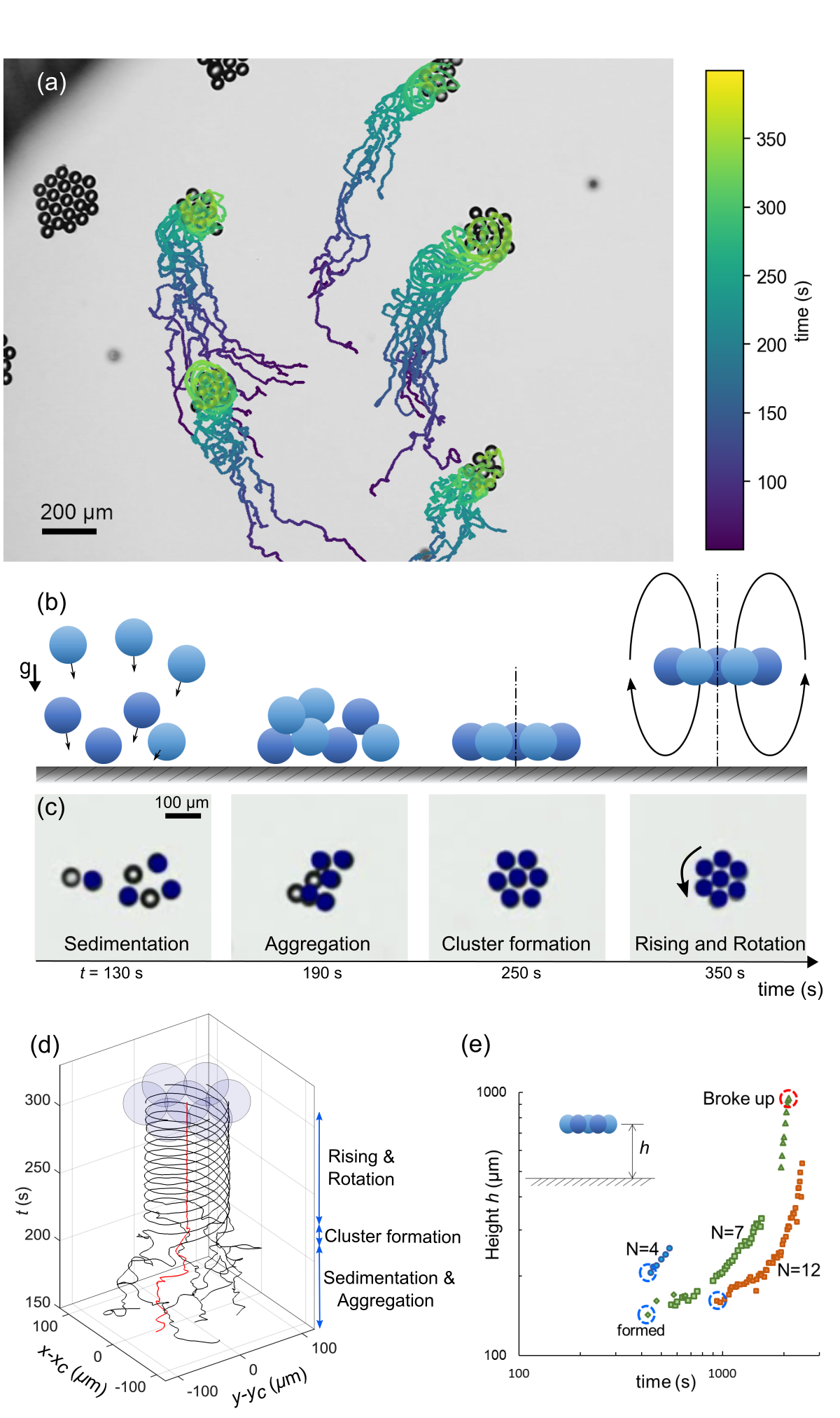}
  \caption{(a)  Flow-induced self-assembly of droplets, resulting in spontaneously-rotating clusters, with trajectories of individual droplets colour coded by time. The micrograph is one snapshot from the experiment. (b) and (c): four stages of rotating cluster formation,  schematic drawings and video micrographs. The coloured droplets in (c) lie in the focal plane of the objective. (d) Trajectories of droplets forming one cluster. The z-axis represents the elapsed time. (e) Evolution of cluster hovering height with time, for clusters of various size.
    \label{fig:clustering} }
\end{figure}

\section{Phase diagram for the cluster dynamics}
We now study the phase behaviour of the clusters and characterise their dynamics. 
The two parameters under investigation are cluster size $N$ and P\'eclet number ($Pe$). We chose $N$ as a simple measure of collectivity in the cluster, 
while $Pe$, being a measure for oscillatory instabilities in single droplets~\cite{hokmabad2021_emergence}, should also be applicable to classify similar instabilities in clustered states.

In a space spanned by droplet activity ($Pe$) and cluster size ($N$), we observe four regimes (\figref[a]{fig:PhaseDiagram} and Movie S5): (i) No clustering due to insufficient droplet activity, (ii) Non-rotating clusters, (iii) Rotating clusters, and (iv) unstable clusters due to high activity. 

At low $Pe$, droplets generate weaker flow and only small clusters form ($N=3-5$ at $Pe=5.2$). By increasing $Pe$, the droplet activity increases and in turn the hydrodynamic interactions are strong enough to form larger clusters.

At $Pe=6.8$ to 8.7, the dynamics of smaller clusters ($N<8$) transition from non-rotating to rotating. 

For sufficiently large $Pe$, we observe unstable clusters that rearrange their structure frequently. 
In this regime, during the very short periods when the cluster is hexagonally ordered, it starts to rotate before reconfiguring again into a disordered aggregate.

While increasing $Pe$ results in the formation of larger clusters, the cluster size also influences the collective dynamics. As shown in \figref[a]{fig:PhaseDiagram}, at a given $Pe$, the cluster behaviour changes from rotating to non-rotating with $N$. For instance, at $Pe=10.9$, very large clusters of size $N>20$ barely rotate, while smaller clusters ($N<10$) rotate steadily.
This is reflected in \figref[b]{fig:PhaseDiagram}, in which we have plotted the measured rotational period of clusters with different $N$ at $Pe=13.2$. The rotational velocity of a cluster linearly decreases with $N$. 

A similar dependence on $N$ was found by Petroff et al.~\cite{petroff2015_fast-moving}  in analysis of \textit{Thiovulum majus} bacteria that self-organise into 2D clusters rotating counter-clockwise at a solid boundary~\cite{das2019_transition}.
Here, the counter-clockwise orientation is imposed by the chirality of the bacterial flagella, and a global torque results from the coupling of the individual spinners. The authors analysed the dependence of the rotation period on the cluster size using a scaling relation between viscous friction and hydrodynamic torques.

As in our case the symmetry is broken spontaneously, we observe that the clusters rotate both clockwise and counter-clockwise and can spontaneously switch between the two. We illustrate this with sequential video frames in \figref[c]{fig:PhaseDiagram}, taken from  Movie S6, where a cluster of size $N=9$ first steadily rotates clockwise, then spontaneously loses its crystalline order, then again reorders and begins to rotate, but now counter-clockwise.

\begin{figure*}
  \centering
  \noindent\includegraphics[width=0.9\textwidth]{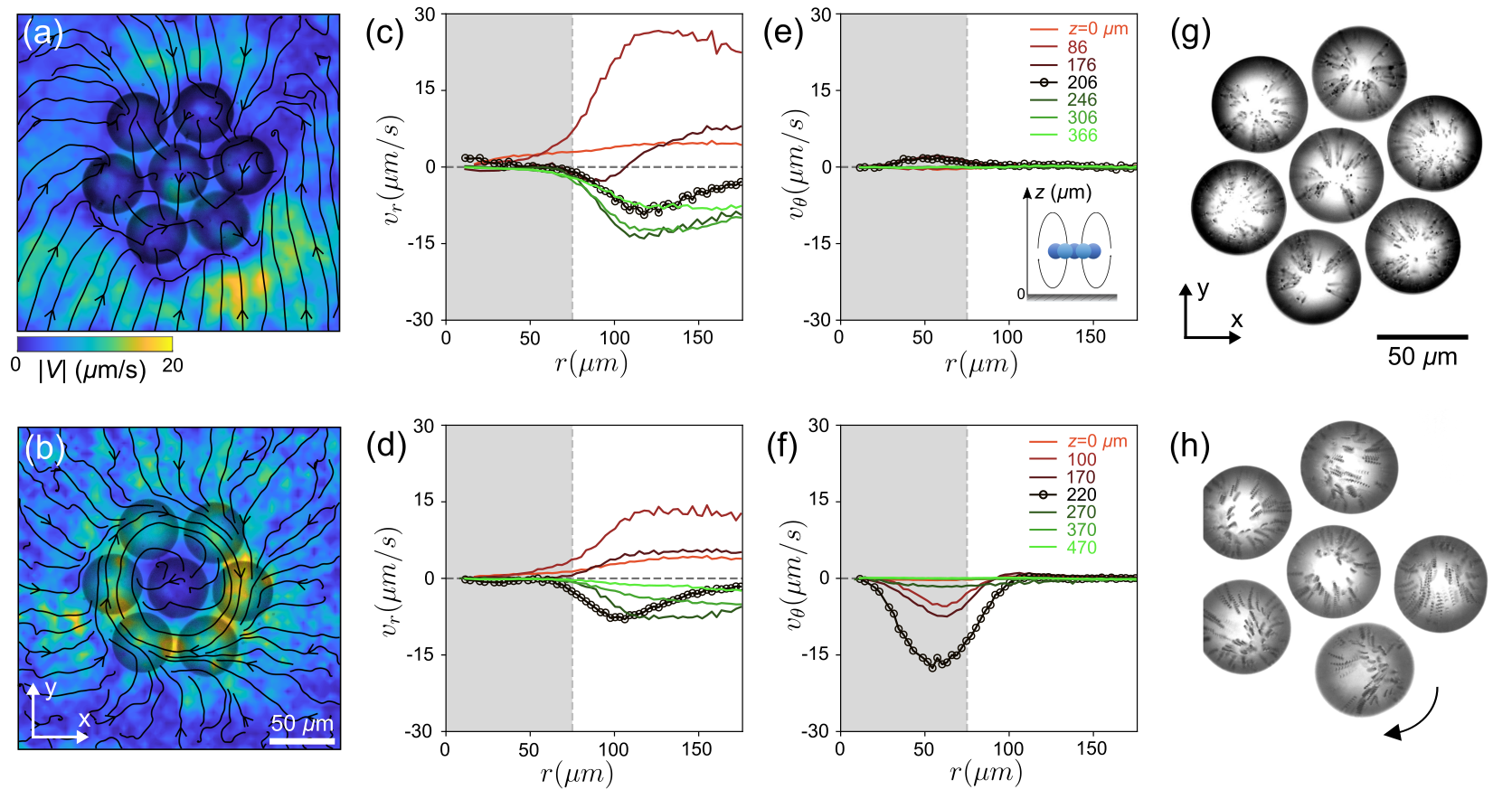}
  \caption{(a) and (b) Flow-fields generated by the clusters in the $x-y$ plane, by non-rotating and rotating clusters, respectively. (c) and (e) Radial $v_r$ and tangential  $v_\theta$ components of the velocity field at different heights for a non-rotating cluster. The black profile with the circle marker corresponds to the height of the cluster plane. (d) and (f) same for a rotating cluster. (g) and (h) internal convective flows within the droplets visualised by streaklines of tracer colloids during 0.25 seconds, for non-rotating and rotating clusters, respectively.
    \label{fig:flowfields} }
\end{figure*}

\begin{figure}
  \centering
  \noindent\includegraphics[width=0.45\textwidth]{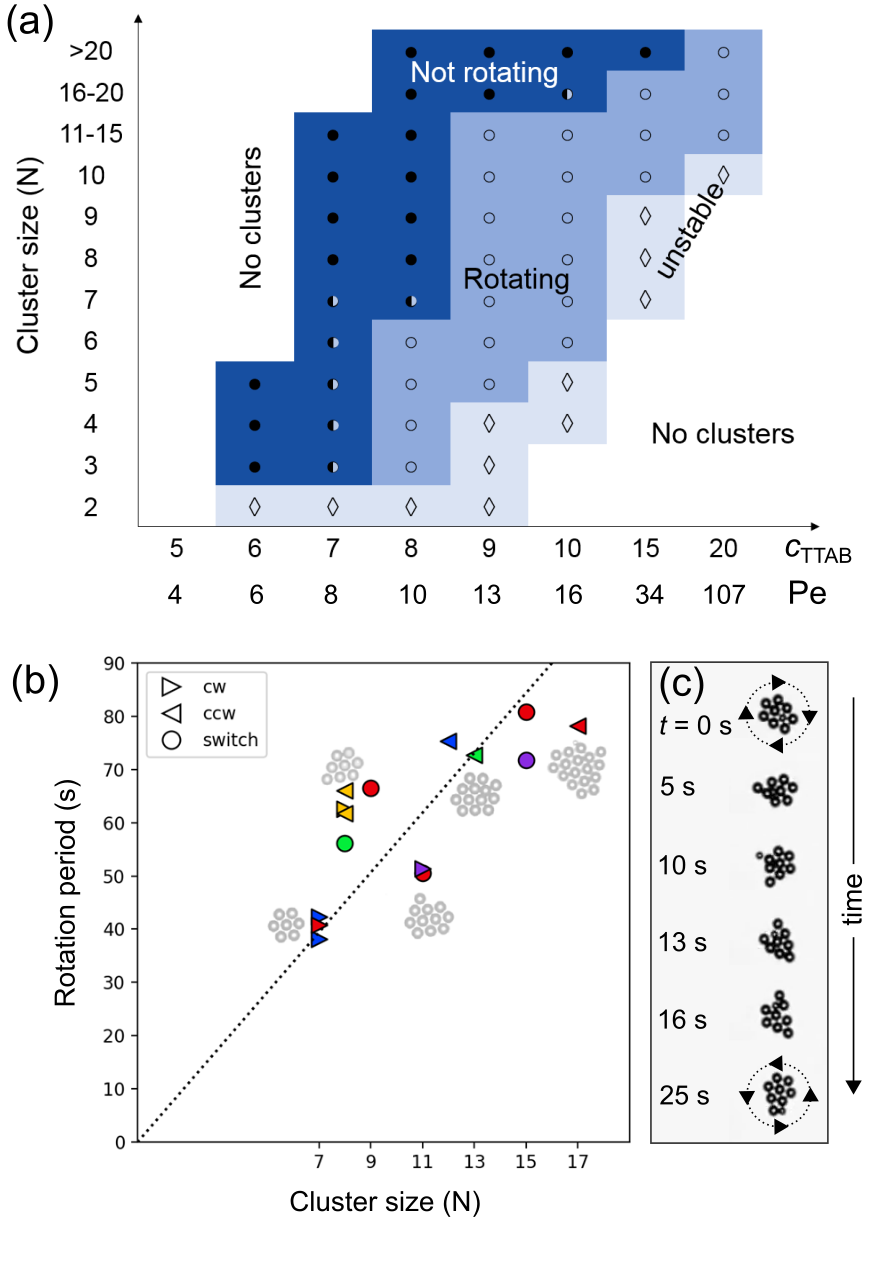}
  \caption{(a) The phase diagram of the cluster dynamics and stability. In the horizontal axis, the $c_{\rm TTAB}$ values and the corresponding $Pe$ values are presented. $\bullet$ and $\circ$ denote non-rotating and rotating clusters, respectively. The half-full circle corresponds to the margin where both behaviours are observed. $\diamond$ corresponds to unstable clusters. (b) Rotational period of the clusters versus their size, labels denote clockwise, counterclockwise and switching rotation. (c) A cluster switching rotation direction after dissociating and reforming again.
    \label{fig:PhaseDiagram}}
\end{figure}

\begin{figure}
    \centering
    \includegraphics[width=\columnwidth]{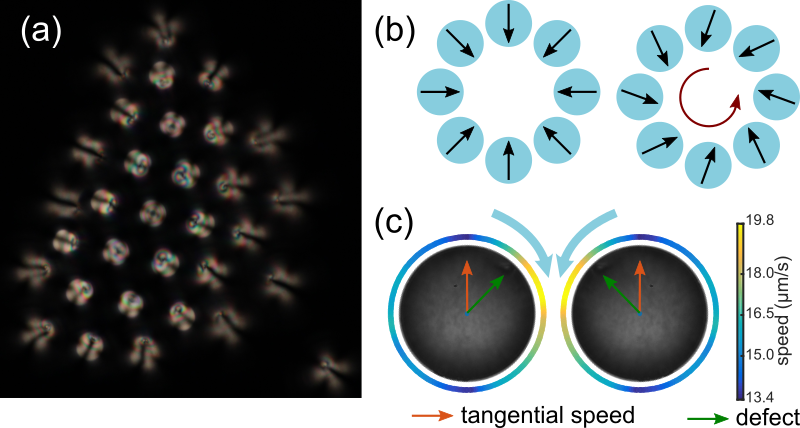}
    \caption{(a) Polarized micrograph of a non-rotating cluster of 5CB nematogen droplets. The orientation of the internal defect is inferred via the fringe pattern: swimmers around the perimeter are titled inwards. (b) Defect orientation pattern required for a non-rotating (left) and a rotating (right) cluster. (c) The coupling of tangential flow fields of adjacent droplets would cause the droplets to tilt towards each other. Panel adapted from~\cite{kruger2016_curling}, Fig. 4}
    \label{fig:5cbclusters}
\end{figure}
\section{Chirality and broken symmetries}
We further note an intriguing fact when we compare the present results to our previous study on clustering droplets composed of CB15's nematic isomer 5CB~\cite{kruger2016_dimensionality}.
Individual \textit{nematic} droplets feature an extremely robust helical or oscillatory instability caused by nematoelastic torques~\cite{kruger2016_curling}, while, as discussed above, helical swimming in \textit{isotropic} droplets is limited to a narrow $Pe$ range and mostly suppressed in quasi 2D. However, for clustered states the situation is reversed: isotropic clusters develop stable rotation states, while, in experiments exploring the same parameter space as in \figref{fig:PhaseDiagram} using 5CB droplets, we did not observe any collective rotation.

To understand why a cluster built from nematogens would not rotate, let us consider the sketches  in \figref{fig:5cbclusters}.

Motile nematic droplets with perpendicular interfacial anchoring incorporate a topological defect of charge $+1$ that is displaced by the internal flow towards the anterior of the moving droplet. This confers a polarity on the droplet, symbolised in \figref{fig:5cbclusters} by a vector corresponding to the defect orientation, which can be extracted from polarised or phase contrast microscopy (panels a and c). As the viscosity of nematogens is reduced around a defect~\cite{chmielewski1986_viscosity}, the tangential flow at the interface is enhanced in its vicinity (panel c).
 In the collective state, the strong radial outflux between cluster and container bottom tilts the droplets around the perimeter inwards (panel a and b, left, see also Fig. 5 in~\cite{kruger2016_dimensionality}). A superimposed rotation would require an additional in-plane tilt of all defect vectors in the same direction (panel b, right). However, as sketched in panel c, the superposition of tangential flow between adjacent droplets would favour them to turn towards each other, such that a uniform tilt causing collective rotation could not arise spontaneously. 

Inspired by systems like \textit{Thiovulum majus}, we can yet enforce a prescribed rotation on the stationary nematic clusters by adding the enantiomeric chiral dopants R811 and S811 to the 5CB oil phase. This results in the formation of helical superstructures inside the droplets and, in turn, a twisting flow around them~\cite{yamamoto2019_hydrodynamic}. The clusters now rotate with the chirality imposed on the molecular scale (\figref[b,c]{fig:LC}). In this case, the chirality is imposed and not required to emerge spontaneously. 

\section{Conclusions}

In conclusion, we demonstrated that a simple emulsion of active droplets can use hydrodynamic interactions to self-organise into rotating planar clusters reminiscent of living systems. 

Since the system has no inherent broken symmetry -- the droplets themselves are isotropic and rotation occurs even in symmetric heptamer structures like the one in \figref[c]{fig:clustering} --  the emergence of this rotational motion was, to us, unexpected. Lacking built-in asymmetries, the symmetry has to be broken spontaneously by the nonlinear dynamics of the system.

This helical instability arises due to the nonlinearity of surfactant transport and circumvents the need to incorporate built-in, structural asymmetry within the droplet~\cite{kruger2016_curling,hokmabad2019_topological,wang2021_active}.  We characterised the 3D dynamics of individual droplets and identified for increasing $Pe$ three regimes of persistent, helical, and chaotic swimming.

Notably, the helical instability in individual CB15 droplets is limited to a very narrow range of $6.8\le Pe\le 8.7$. Beyond $Pe=8.7$, the dynamics become increasingly chaotic. In contrast, the planar clusters exhibit rotational dynamics within the range of $6.8\le Pe\le 91.8$ which is remarkably broader compared to the range for the individual droplets. In other words, the clusters show ordered rotational motion even at a $Pe$ range corresponding to the chaotic motion of its constituent elements. This suggests that the cluster formation suppresses the chaotic dynamics of the individual droplets and integrates them into an ordered collective system which performs a synchronised rotational motion. 
This stability is achieved cooperatively through mutual hydrodynamic interactions between the constituent elements of the clusters. At very high $Pe$, only very large clusters maintain their ordered structure (stability) and rotate.  We showed that the collective properties of rotation period and stability can be controlled by tuning the activity ($Pe$) and the cluster size.

In our comparison with 5CB droplets, we found that the torques that cause stable helical rotation for individual swimmers can inhibit rotation on the collective scale.
We found this quite remarkable: CB15 and 5CB are isomers, so that the main difference between the two systems lies in nematic topology. For individual swimmers, the nematic system has a much more robust helical instability. However, the nonlinear dynamics governing the emergence of collective states apparently cause an inversion of the rotational characteristics, as CB15 clusters rotate and 5CB ones don't, a phenomenon that can be rationalised by fundamental symmetry arguments. Combined with the possibility of controlling chirality via chiral dopants, our findings provide  inspiration to the design of tailored rotational states in functional materials.

\begin{figure}
  \centering
  \noindent\includegraphics[width=0.35\textwidth]{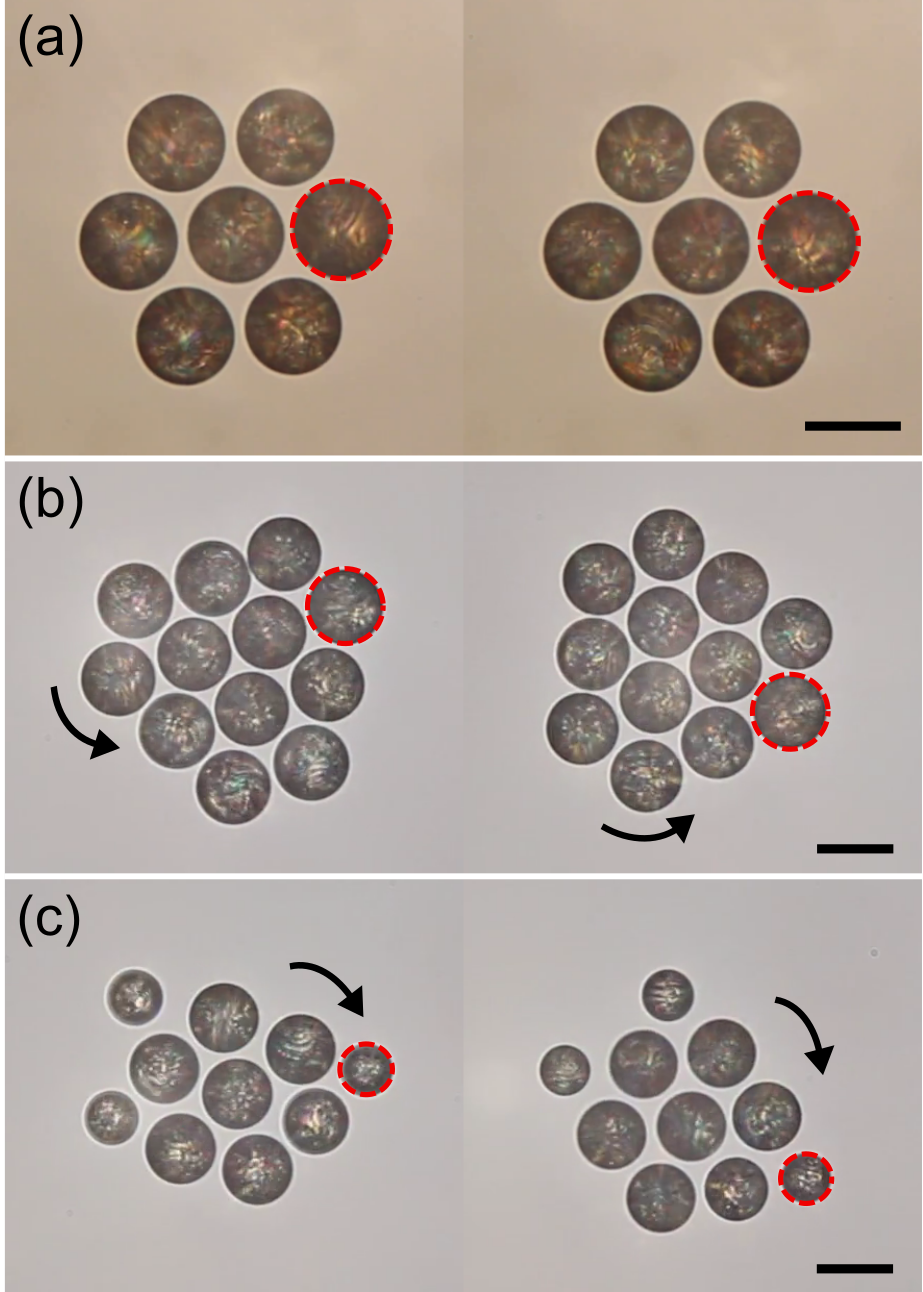}
   \caption{The behaviour of clusters formed by active (a) nematic 5CB, (b) chiral nematic 5CB+R811, and (c) chiral nematic 5CB+S811 droplets. The dashed circle shows one selected droplet within each cluster between two sequential frames (25 seconds apart).
    \label{fig:LC}}
\end{figure}

\section*{Author Contributions}
B.V.H. and A.N. designed and performed experiments, analyzed data and wrote the manuscript. P.R. analyzed data and wrote the manuscript, C.C.M. designed experiments, analyzed data and wrote the manuscript.

\section*{Conflicts of interest}
There are no conflicts to declare.

\section*{Acknowledgements}
We acknowledge fruitful discussions and advice specifically from Ranabir Dey, as well as with Jun Hayakawa, Detlef Lohse, Yibo Chen and Stephan Herminghaus. C.C.M. and B.V.H. acknowledge financial support from the DFG Priority Programme 1726, "Microswimmers". The authors acknowledge financial assistance from Hitachi Ltd.

\section{Appendix: Materials, methods and supplementary movies}\label{sec:clAppx}

\subsection{Materials}
We obtained 5CB, CB15, TTAB, R811, and S811 from commercial suppliers (Synthon Chemicals and Sigma-Aldrich) and used them as is.

\subsection{Droplet production}
We fabricated microfluidic devices using standard soft
lithography procedures and used them to produce monodisperse emulsions of CB15 and stored them in aqueous TTAB solution with a submicellar concentration. For a detailed protocols of the droplet production, see Ref. ~\cite{hokmabad2021_emergence}.

\subsection{Experimental cells and protocols}
For the quasi-2D geometries used for PIV measurements, we embedded rectangular reservoirs with the dimensions of  $13\times8\times \SI{0.05}{\mm^3}$ via UV photolithography in SU-8 photoresist (Micro Resist Technologies). For a detailed description of the fabrication protocol, refer to the supporting information in  ref.~\cite{hokmabad2021_emergence}. 

For 3D experiments, we used a cylindrical PDMS reservoir (\figref{fig:setup}) of  height 5 mm and diameter 5 mm. We added small amounts of stock emulsion to TTAB solution at the target concentration, filled reservoirs with this mixture, capped the reservoir with glass cover slips and started videomicroscopy as fast as experimentally feasible (within tens of seconds). 

\subsection{Microscopy, image recording, and analysis}
We observed the large-scale dynamics of the active emulsions under a bright-field microscope (Leica DM4000 B) at magnifications of 2.5x and 5x. Video frames were recorded at 4 frames per second using a Canon digital camera (EOS 600d) with a digital resolution of 1920 x 1080 px. The droplet coordinates were extracted using custom Python scripts (available on request) based on the libraries numpy, PIL, and openCV. Droplet coordinates were acquired via a sequence of background correction, binarisation, feature correlation, blob detection by contour analysis, and minimum enclosing circle fits. Droplet trajectories were then acquired via a frame-by-frame nearest-neighbour analysis. 
The height measurements and PIV measurements were performed using Olympus IX83 microscope and images were recorded by a 4 MP CMOS camera (FLIR Grasshopper 3, GS3-U3-41C6M-C) at frame rates between 4 and 24 frames per second. PIV analysis was done using the MATLAB-based  PIVlab interface~\cite{thielicke2014_pivlab}.

\subsection{Viscosity measurements}
We measured the viscosity of the oil phase and the surfactant solution using an Anton Paar MCR 502 rotational rheometer. For the oil phase, we found $\mu_{\rm i}=138\, \rm mPa.s$,  for the surfactant solutions the measured values were $\mu_{\rm i}=$ 1.3, 1.6, 1.9, and 2.9 mPa.s, corresponding to $c_{\rm TTAB}=$ 5, 10, 15, and 20 $\rm wt.\%$, respectively. The estimated values, corresponding to other $c_{\rm TTAB}$, were extracted by interpolation. 

\subsection{Dissolution rate of active droplets}
\begin{figure}
  \centering
  \noindent\includegraphics[width=0.3\textwidth]{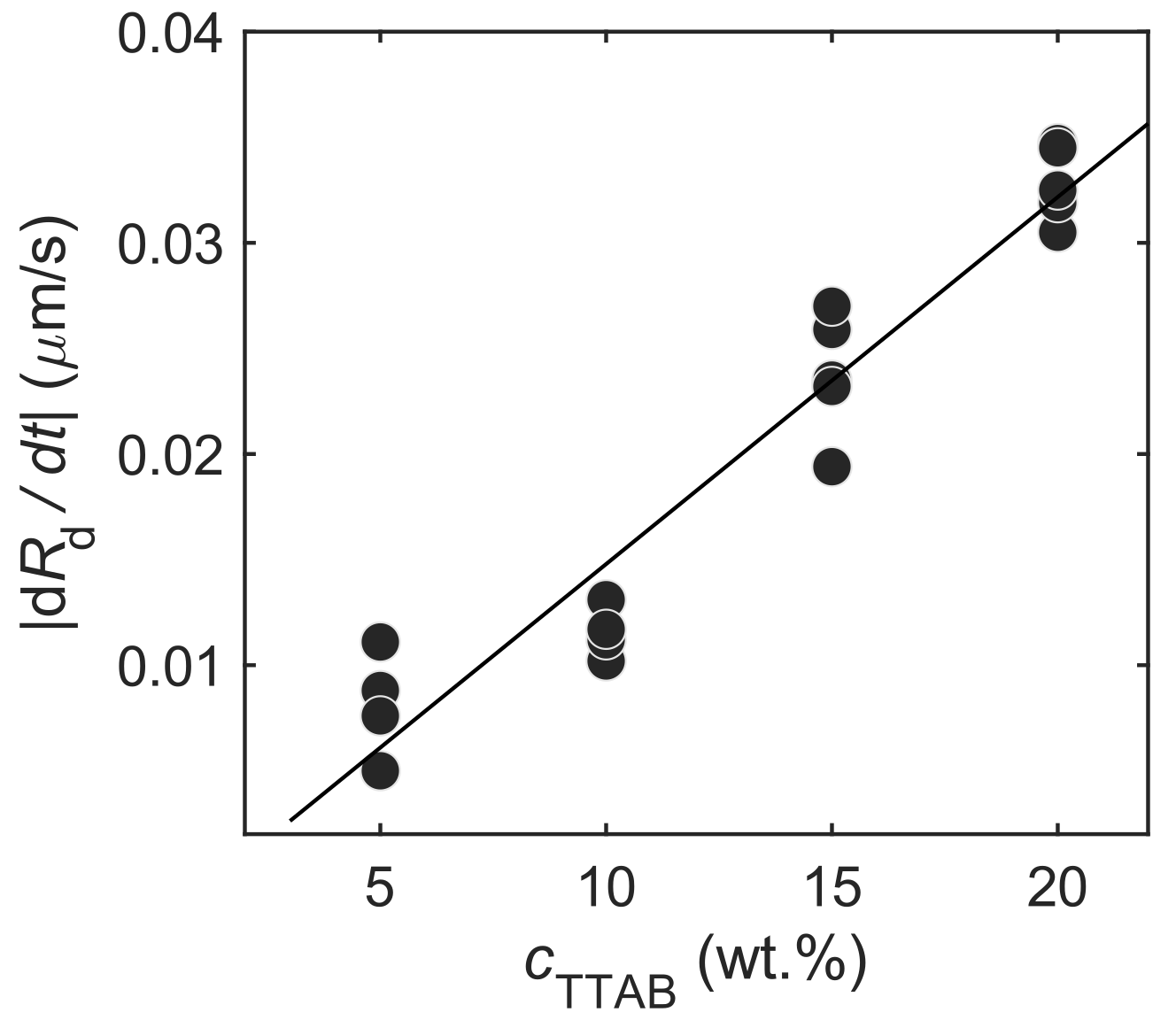}
  \caption{Dissolution rate of active droplets ${\rm d}R_d/{\rm d}t$ versus different $c_{\rm TTAB}$ values. 
    \label{fig:SolRate} }
\end{figure}

To estimate the surfactant consumption rate $q_s$ for our calculation of $Pe$, we measured the droplet shrinking rate ${\rm d}R_d/{\rm d}t$ (plotted in \figref{fig:SolRate}). We found a dependence on $c_{\rm TTAB}$. We used a first order approximation, via linear regression (black line), for ${\rm d}R_d/{\rm d}t$, to calculate $q_s$ for different $c_{\rm TTAB}$ values. $q_s$ is estimated by relating the dissolution rate of the active droplet to the isotropic surfactant consumption at the droplet interface~\cite{izri2014_self-propulsion}

\subsection{Brinkman squirmer model}
\label{app_Brinkman}
Following the formalism in ref.~\cite{jin2021_collective}, we consider a squirmer of radius $r=R$ moving through a Hele-Shaw cell of height $h$ at low Reynolds number. The Stokes equations approximated by the Brinkman equations can be written as
    \begin{equation}
        \vec{\nabla} p = \mu (\nabla^2 - k) \vec{u}, \quad \vec{\nabla} \cdot \vec{u} = 0,
    \end{equation}
where $k = 12/h^2$ is the permeability. Assuming that the interface is impermeable and the flow field is determined by the tangential velocity, the boundary conditions are
\begin{equation}
u_{r}^n(r=R)=0  
\end{equation}
\begin{equation}
u_{\theta}^n(r=R)=-\dfrac{R}{h}\dfrac{K_{n-1}(R/h)}{K_{n}(R/h)}b_n \sin(n\theta)
\end{equation}
where $K_{n}(x)$ is the modified Bessel function of the second kind of order $n$.
\\
The solution of the above equations is given below.
\\
The surface actuation solution using the first two modes is:
\begin{equation}
    {u_{r}}^{act}=\sum_{n=1}^{2} b_n \left[ n \left(\frac{R}{r}\right)^{n+1} -n\frac{R}{r}\frac{K_n(r/h)}{K_n(R/h)} \right]\cos(n\theta)
\end{equation}
\begin{equation}
    {u_{\theta}}^{act}=\sum_{n=1}^{2} b_n \left[ n \left(\frac{R}{r}\right)^{n+1} +\frac{R}{h}\frac{K'_n(r/h)}{K_n(R/h)} \right]\sin(n\theta)
\end{equation}
\begin{equation}
    U_{act}={u_{r}}^{act}+{u_{\theta}}^{act}
\end{equation}
The translation solution is:

\begin{equation}
        {u_{r}}^{tr} =  U \left[ \frac{R^2 K_2\left(\frac{R}{h }\right)-2 h  r K_1\left(\frac{r}{h }\right)}{r^2 K_0\left(\frac{R}{h }\right)} \right] \cos (\theta ),
\end{equation}
\begin{equation}
        {u_{\theta}}^{tr} = U \left[ \frac{R^2 K_2\left(\frac{R}{h }\right)
        -2 h  r K_1\left(\frac{r}{h }\right)
        -2 r^2 K_0\left(\frac{r}{h }\right)
        }{r^2 K_0\left(\frac{R}{h }\right)} \right] \sin (\theta )
\end{equation}      
\begin{equation}
    U_{tr}={u_{r}}^{tr}+{u_{\theta}}^{tr}
\end{equation}    
The complete solution is the sum of surface actuation and translation solutions,
\begin{equation}
    U_{swim}=U_{act}+U_{tr}
\end{equation}   
Then, the experimental flow field $U_{exp}$ measured by particle image velocimetry is used to minimise the residue $U_{exp} - U_{swim}$ to determine the coefficients $b_1$ and $b_2$.

\subsection{Supplementary movies captions}

\textbf{Movie S1.} Hydrodynamic interaction between two active droplets within a Hele-Shaw cell. The video is played $6\times$ faster. The background shows a multi-frame superposition tracking the trajectories of the two droplets.

\textbf{Movie S2.} Phase separation and cluster formation in an active emulsion inside a deep reservoir, including both rotating and non-rotating clusters. The video is played $24\times$ faster.

\textbf{Movie S3.} Schematic rendering (Blender) of the assembly and collective motion of a rising and rotating cluster. The $xy$ data was taken from bright field microscopy data of $\SI{50}{\um}$ droplets, while the height was assumed to increase linearly with time. 
The video is played $8\times$ faster.

\textbf{Movie S4.} Internal convective flows within the droplets visualised by streaklines of tracer colloids for non-rotating and rotating clusters. The video is played in real time.

\textbf{Movie S5.} Different regimes of cluster dynamics. The video is played $24\times$ faster.

\textbf{Movie S6.} A cluster switching rotation direction after dissociating and reforming again. The video is played $24\times$ faster.

%

\end{document}